\documentclass[11pt]{article}
\pdfoutput=1
\usepackage{cite}
\usepackage{graphicx}

\usepackage{amsmath,amsfonts}
\usepackage{hyperref}
\usepackage{comment}
\usepackage{fullpage}
\usepackage[ruled,vlined, linesnumbered]{algorithm2e}

\newcommand*{\email}[1]{%
    \normalsize\href{mailto:#1}{#1}\par
    }

\begin{document}
\title{R(QPS-SERENA) and R(QPS-SERENADE): Two Novel Augmenting-Path Based Algorithms for Computing Approximate Maximum Weight Matching}
\author{%
\begin{tabular}{cc}
Long Gong & Jun (Jim) Xu \tabularnewline
\email{gonglong@gatech.edu} & \email{jx@cc.gatech.edu} \tabularnewline
\multicolumn{2}{c}{Georgia Institute of Technology, USA}\tabularnewline
\end{tabular}
}

\maketitle 

\begin{abstract}
In this addendum, we show that the switching algorithm QPS-SERENA can be converted R(QPS-SERENA), an algorithm for computing approximate Maximum Weight Matching (MWM).
Empirically, R(QPS-SERENA) computes $(1-\epsilon)$-MWM within linear time (with respect to the number of edges $N^2$) for any fixed $\epsilon\in (0,1)$, for complete bipartite graphs
with {\it i.i.d.} uniform edge weight distributions. This efficacy
matches that of the state-of-art solution, although we so far cannot prove any theoretical guarantees 
on the time complexities needed to attain a certain approximation ratio. 
Then, we have similarly converted QPS-SERENADE to R(QPS-SERENADE), which empirically should
output $(1-\epsilon)$-MWM within only $O(N \log N)$ time for the same type of complete bipartite graphs as described above.
\end{abstract}

\section{Introduction}\label{sec: intro}

To algorithmically find a matching whose weight is largest among all matchings over a weighted bipartite or general graph, known as maximum weight matching (MWM) computation,
is a well-studied research problem in combinatorial optimization.  Computing MWM has so many real-life applications that research efforts towards such efficient algorithms  
started more than $150$ years ago by Jacobi.  For example, a classical application is the assignment problem
~\cite{kuhn1955hungarian,kuhn1956variants}, in which $n$ workers and $n$ jobs are viewed as the two disjoint vertex sets of 
a complete bipartite graph, and the edge between worker $i$ and job $j$ is weighted by worker $i's$ competence score in performing job $j$;  in this problem setting a MWM corresponds to a
workers-to-jobs assignment that maximizes the total competence score.  
To this day, the most efficient algorithm for computing a MWM over  
a general graph with arbitrary nonnegative weights
has a time complexity of $O(EV + V^2\log V)$~\cite{Gabow1990fastestMWM}, where $E$ and 
$V$ are the number of edges and vertices in the graph respectively.  While this algorithm is efficient enough for 
most applications where either $V$ is small (say no more than tens of thousands) or the graph is sparse ({\it i.e.,} $E = O(V)$),
it may not be for others that have a massive (in $V$) dense underlying graph, such as the graph partitioning problem in 
VLSI Design~\cite{Monien2000}.  


For most applications, finding an approximate MWM is often good enough.   In this case we would like to compute a $\delta$-MWM, which is 
a matching whose weight is at least $\delta$ ($0< \delta < 1$) times that of MWM.  The algorithmic problem of computing approximate MWM also has a long history. 
Over these years, many linear-time (with respect to $E$ given a fixed $\delta$) or near-linear-time algorithms ({\it e.g.,} \cite{Duan2014LinearAppMWM,Duan2010LinearAppMWM,Pettie2004LinearAppMWM,Meinel1563LinearAppMWM}) 
have been proposed for computing approximate MWM; a linear-time algorithm is clearly optimal with respect to $E$, since just to read in all the edge weights takes $O(E)$ time.
Among them, the linear-time algorithm proposed in~\cite{Duan2014LinearAppMWM} achieves the best approximation ratio with respect to $\delta$. In particular, 
it is a deterministic algorithm that guarantees to output, for any (tiny) $\epsilon > 0$, a $(1-\epsilon)$-MWM in $O(E\epsilon^{-1}\log \epsilon^{-1})$ time.


We report two related inventions in this addendum.  Our first invention is to convert QPS-SERENA, described in~\cite{Gong2017QPS}, 
into an approximate MWM algorithm.  We call the latter algorithm R(QPS-SERENA), since it applies QPS-SERENA repeatedly. 
Our second invention is to similarly convert QPS-SERENADE, the parallelized QPS-SERENA, into a parallel approximate MWM algorithm
which we call R(QPS-SERENADE).  Here QPS-SERENADE is the augmentation of E-SERENADE~\cite{GongLiuYangEtAl2018} using QPS~\cite{Gong2017QPS} 
in the same way as that of SERENA.  We emphasize that the role of QPS is 
essential in both R(QPS-SERENA) and R(QPS-SERENADE):  Neither R(SERENA) nor R(SERENADE) exists because both SERENA and SERENADE requires packet arrivals in each ``time slot'' to operate, which 
do not exist after the conversion.

We show that, given any fixed $\epsilon > 0.02$, the R(QPS-SERENA) algorithm can (empirically) output 
a $(1-\epsilon)$-MWM in $O(E)$ time on average over a set of complete bipartite graphs with {\it i.i.d.} uniform random edge weights. 
Note this time complexity already matches that of the aforementioned state-of-art solution~\cite{Duan2014LinearAppMWM}
for computing approximate MWM under dense bipartite graph. 
More exiting news is that  
R(QPS-SERENADE), 
can empirically do the same in $O(V\log V)$ time. This 
R(QPS-SERENADE) 
result is potentially ground-breaking
because as mentioned in~\cite{GongLiuYangEtAl2018}, 
MWM computation (exact or approximate) is notoriously hard to parallelize: 
All existing parallel and distributed approximate MWM solutions, except one based on the approach of Belief Propagation (BP)~\cite{Bayati2008BP,Bayati2005BP}, require at least $O(E)$ processors, which is not practical for large dense graphs 
(where $E = O(V^2)$),
whereas R(QPS-SERENADE) needs only $O(V)$ processors or less.  The BP-based solution~\cite{Bayati2008BP,Bayati2005BP} computes an exact MWM using a total of $O(V^3)$ computation time evenly distributed across $O(V)$ processors,
resulting a per-processor time complexity of $O(V^2)$ that is larger than needed by R(QPS-SERENADE) empirically.


\section{Our Results}\label{sec: our-results}

We describe in Section~\ref{subsec: r-qps-serena} how to convert QPS-SERENA~\cite{Gong2017QPS}, an online crossbar scheduling algorithm, to 
R(QPS-SERENA), an offline matching algorithm.  We then explain in Section~\ref{subsec: r-qps-serenade} how QPS-SERENADE~\cite{GongLiuYangEtAl2018} is similarly converted 
to R(QPS-SERENADE).  Before we do both, in Section~\ref{subsec: problem-formulation} we formulate the approximate MWM problem in a bipartite graph and explain how an offline matching problem is reduced to an online switching problem.



\subsection{Problem Formulation and Reduction}\label{subsec: problem-formulation}

Consider a weighted bipartite graph $G(U, V, E, w)$, where $U$ and $V$ are the two independent
vertex sets, $E$ is the set of edges, and every edge $e \in E$ is assigned a weight value $w(e)$.
We assume that $|U| = |V| = N$, {\it i.e.,} the bipartite graph is balanced.   This assumption is not restrictive 
because any unbalanced bipartite graph can be converted to a balanced one by adding 
dummy vertices to the independent vertex set that contains less vertices, and if necessary dummy edges with weight $0$.
We also assume 
that all edges have nonnegative weights, because edges with negative weights can be safely
ignored:  Adding them to a matching would only decrease the weight of the matching.
Our computation problem is, given any weighted balanced bipartite graph $G(U, V, E, w)$, to compute 
an approximate MWM.  

In both R(QPS-SERENA) and R(QPS-SERENADE), the first step is to reduce this matching computation problem to a crossbar scheduling problem as follows.
Given any weighted balanced bipartite graph $G(U, V, E, w)$ (as the input to this matching computation problem), 
we map it to an $N\times N$ input-queued crossbar switch, and map 
vertex sets $U$ to input ports and $V$ to output ports respectively.   Each edge is correspondingly mapped to a VOQ
as follows.
Suppose a vertex $u \in U$ is mapped to input port $i$, and a vertex $v \in V$ is mapped to output port $j$.  Then the edge 
$e = (u, v)$ is mapped to the VOQ at input port $i$ that buffers packets destined for output port $j$, and the length of this VOQ 
is considered to be $w(e)$.  If there is no edge between two vertices, then the corresponding VOQ is assumed to have length $0$.
The resulting crossbar scheduling instance ({\it i.e.,} the weights of these
$N^2$ VOQs) will be used as the input to our switching algorithms R(QPS-SERENA) and R(QPS-SERENADE), to ``fool" them into
eventually outputting an approximate MWM, as we will explain next.

\subsection{R(QPS-SERENA)}\label{subsec: r-qps-serena}

R(QPS-SERENA) is simply to run
QPS-SERENA~\cite{Gong2017QPS} repeatedly over multiple time slots as follows.
During each time slot, the crossbar scheduling instance described above 
is used as the input to (the QPS part of) the QPS-SERENA algorithm for computing
the crossbar schedule for the time slot.  However, unlike in the normal operations of
QPS-SERENA for crossbar scheduling, where the lengths of 
the VOQs need to be updated according to the packet departures dictated by the 
computed crossbar schedule and packets arrivals during the current time slot, 
in R(QPS-SERENA) there are no packet arrivals
and departures during any time slot.  In other words, the same crossbar scheduling instance 
is fed to QPS-SERENA over and over, to ``fool" it into computing an approximate MWM with 
respect to the corresponding weighted bipartite graph.

As explained in~\cite{Gong2017QPS}, the 
matching (crossbar schedule) output by QPS-SERENA either stays the same or grows larger in weight
one time slot after another, and hopefully becomes an approximate (or exact) MWM after 
a small number of time slots.  Indeed, we will show in Section~\ref{sec: performance} that, after only $O(N)$ time slots, empirically the resulting matching 
is an approximate MWM with high probability for a certain family of weighted bipartite graphs.  

Now we emphasize a subtle difference between this crossbar scheduling problem, reduced from 
the approximate MWM computation problem, and that in the normal context of 
switching.  The former problem is offline in that any algorithm solution to it is required 
to output only the final matching (for the ``last time slot") at the end;  the solution does not have to 
output the crossbar schedule
for each time slot in real-time ({\it i.e,} at the beginning of the time slot), which is 
required in the latter problem that is online.  
The offline nature of the former problem allows for a host of algorithmic tricks that cannot be used 
in a solution to the latter (online) problem.  

As shown in~\cite{giaccone2003randomized}, the time complexity of SERENA alone is $O(N)$.
The time complexity of QPS alone is $O(1)$ per port
(or $O(N)$ for all $N$ 
ports) when the VOQ lengths are nonnegative integers, as shown in~\cite{Gong2017QPS}, but 
becomes $O(\log N)$ per port ({\it e.g.,} using a standard binary-based-tree implementation) 
when the VOQ lengths are nonnegative real numbers.
However, if R(QPS-SERENA) executes QPS-SERENA at least $O(N)$ times, 
these $O(N)$ QPS computations can be batched together to run in $O(N^2)$ time,
or $O(N)$ time per QPS computation, as we will show in Appendix~\ref{subsubsec: constant-time-qps}.  Note that this batching is 
allowed because, as explained above, R(QPS-SERENA) is dealing with an offline computation here.


\subsection{R(QPS-SERENADE)}\label{subsec: r-qps-serenade}

QPS-SERENADE is the same as QPS-SERENA except that the MERGE operation is performed using E-SERENADE~\cite{GongLiuYangEtAl2018} instead of SERENA.
R(QPS-SERENADE) runs QPS-SERENADE repeatedly and, like in R(QPS-SERENA), every QPS-SERENADE run takes as input the 
same crossbar scheduling instance reduced from the approximate MWM instance.   The intention here of replacing QPS-SERENA by QPS-SERENADE
is to reduce the ``per-time-slot" time complexity from $O(N)$ to $O(\log N)$, so that R(QPS-SERENADE) can produce the same final output as 
R(QPS-SERENA), but in much less time.  For example, if $N$ repetitions are used in both R(QPS-SERENADE) and R(QPS-SERENA), the former has
time complexity $O(N \log N)$ (per port) whereas the latter has time complexity $O(N^2)$.

Recall that the time complexity of QPS-SERENADE
is strictly $O(\log N)$ except that the POPULATE step (for populating a QPS-generated matching that is in general partial into a full 
matching) of SERENADE is ``$O(N)$ light'' ({\it i.e.,} $O(N)$ with a very small constant factor such as $\frac{1}{64}$),
using the bit-parallelism solution proposed in~\cite{GongLiuYangEtAl2018}.   This notion of ``$O(N)$ light'' is acceptable in the switching context, 
because it is effectively $O(\log N)$ for $N$ that is not very large (say $N \le 1,024$).  However, in this theoretical context of approximate MWM computation, 
this notion no longer makes sense.  Fortunately,  
we have obtained the following result recently: 
a distributed iterative algorithm that exactly emulates 
POPULATE in which each port does only $O(\log N)$ work in the worst case.  We will provide brief description of this algorithm 
in Appendix~\ref{subsubsec: paral-population}.

\begin{figure}[!htb]
\centering 
\includegraphics[width=0.9\textwidth]{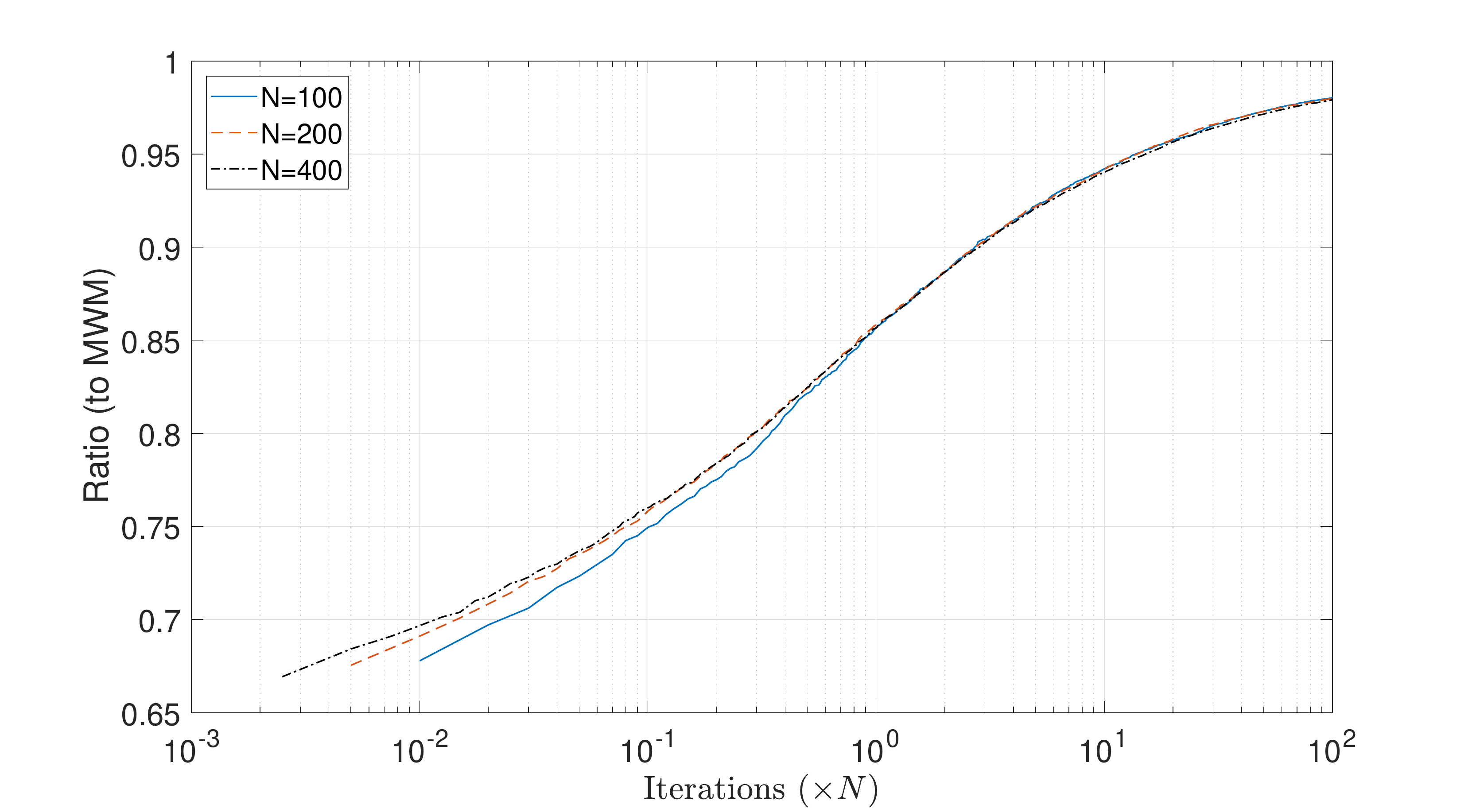}
\caption{Simulation results on complete bipartite graphs.} 
\label{fig: performance-complete-bipartite}
\end{figure}

\subsection{Empirical Performance}\label{sec: performance}

In this section, we evaluate the empirical efficacy of R(QPS-SERENA) in computing approximate MWM.
Note there is no need to evaluate the efficacy of 
R(QPS-SERENADE), since it exactly emulates R(QPS-SERENA). 
We run R(QPS-SERENA) to compute approximate MWM matchings over  
$3$ weighted complete balanced bipartite graphs.  In these three graphs, 
the parameter $N$ takes values $100$, $200$, and $400$ respectively;  in each graph, 
the weight of every edge is an {\it i.i.d.} random variable uniformly distributed 
over the interval $[10.0, 100.0]$.
    

For each of these three graphs, \autoref{fig: performance-complete-bipartite} plots, on the $y$-axis, the weight of the matching output by R(QPS-SERENA) at the end of time (slot) $t$, as a percentage of that of MWM ({\it i.e.,} the approximation ratio),
and plots, on the $x$-axis on a logarithmic scale, time $t$ in the multiple of $N$ time slots.   
For example, for the bipartite graph with $N = 200$, the value $10^0$ ($=1$) on x-axis means time (slot) $1 * N = 200$. 
For each graph, we run R(QPS-SERENA), a randomized algorithm, $100$ times, and each point on the corresponding curve in~\autoref{fig: performance-complete-bipartite} is the average of the weight numbers obtained in these $100$ runs.
\autoref{fig: performance-complete-bipartite} shows that, when the time (in the multiple of $N$ time slots) is large enough 
(say $\geq 1 * N$), these three curves almost overlap with each other. 
That is, for these three graphs, the matchings output by R(QPS-SERENA) at any time slot $t' N$ reach roughly the same approximation ratio relative to their respective MWM's.
For example, for each of the three graphs, after $N$ time slots, R(QPS-SERENA) outputs a matching whose average weight is roughly $0.85$ of that of the respective MWM. 
These experimental results show that, given any fixed approximation ratio, the number of time slots needed for R(QPS-SERENA) to output a 
matching that attains the approximation ratio scales linearly with $N$.  In other words, empirically it takes R(QPS-SERENA) $O(N)$ time slots to reach any fixed approximation
ratio (that is not too close to 1).  Since, for each time slot, the time complexity of QPS-SERENA is $O(N)$, the asymptotic time complexity of R(QPS-SERENA) in this case is $O(N^2)$.


Note this empirical result ($O(N^2)$ computation to reach any fixed approximation ratio) already matches the recent state-of-art 
solution~\cite{Duan2014LinearAppMWM}.  This is exciting news since R(QPS-SERENA) 
is a much simpler algorithm than that in~\cite{Duan2014LinearAppMWM}. 
The more exciting news is that R(QPS-SERENADE), the parallel version of R(QPS-SERENA), can do the same using $N$ processors in 
$O(N \log N)$ time, since its time complexity is $O(\log N)$ per time slot.
This empirical efficacy is better than that of the state of art solution~\cite{WangMahoneyMohanEtAl2015ParallelPackingLP}, 
which can compute a $(1-\epsilon)$-MWM for any dense bipartite graph in $O(\frac{\log^2 N \log \epsilon^{-1}}{\epsilon^2})$ 
with $O(N^2)$ processors.

\section{Conclusion}\label{sec: con}

In this addendum, we show that the switching algorithm QPS-SERENA can be converted R(QPS-SERENA), an algorithm for computing approximate MWM.
Empirically, it outputs $(1-\epsilon)$-MWM within linear time (with respect to the number of edges $N^2$) for any fixed $\epsilon\in (0,1)$, for complete bipartite graphs
with {\it i.i.d.} uniform edge weight distributions.  This efficacy
matches that of the state-of-art solution, although we so far cannot prove any theoretical guarantees 
on the time complexities needed to attain a certain approximation ratio. 
We have similarly converted QPS-SERENADE to R(QPS-SERENADE), which empirically should
output $(1-\epsilon)$-MWM within only $O(N \log N)$ time for the same type of complete bipartite graphs as described above.

\bibliographystyle{IEEEtran}
\bibliography{bibs/app-mwm,bibs/load_balancing,bibs/serenade-references}

\begin{thebibliography}{10}
\providecommand{\url}[1]{#1}
\csname url@samestyle\endcsname
\providecommand{\newblock}{\relax}
\providecommand{\bibinfo}[2]{#2}
\providecommand{\BIBentrySTDinterwordspacing}{\spaceskip=0pt\relax}
\providecommand{\BIBentryALTinterwordstretchfactor}{4}
\providecommand{\BIBentryALTinterwordspacing}{\spaceskip=\fontdimen2\font plus
\BIBentryALTinterwordstretchfactor\fontdimen3\font minus
  \fontdimen4\font\relax}
\providecommand{\BIBforeignlanguage}[2]{{%
\expandafter\ifx\csname l@#1\endcsname\relax
\typeout{** WARNING: IEEEtran.bst: No hyphenation pattern has been}%
\typeout{** loaded for the language `#1'. Using the pattern for}%
\typeout{** the default language instead.}%
\else
\language=\csname l@#1\endcsname
\fi
#2}}
\providecommand{\BIBdecl}{\relax}
\BIBdecl

\bibitem{kuhn1955hungarian}
H.~W. Kuhn, ``The hungarian method for the assignment problem,'' \emph{Naval
  Research Logistics (NRL)}, vol.~2, no. 1-2, pp. 83--97, 1955.

\bibitem{kuhn1956variants}
------, ``Variants of the hungarian method for assignment problems,''
  \emph{Naval Research Logistics (NRL)}, vol.~3, no.~4, pp. 253--258, 1956.

\bibitem{Gabow1990fastestMWM}
\BIBentryALTinterwordspacing
H.~N. Gabow, ``{Data structures for weighted matching and nearest common
  ancestors with linking},'' in \emph{Proc. of the ACM/SIAM SODA 1990}.\hskip
  1em plus 0.5em minus 0.4em\relax Society for Industrial and Applied
  Mathematics, 1990, p. 523. [Online]. Available:
  \url{https://dl.acm.org/citation.cfm?id=320229}
\BIBentrySTDinterwordspacing

\bibitem{Monien2000}
\BIBentryALTinterwordspacing
B.~Monien, R.~Preis, and R.~Diekmann, ``{Quality matching and local improvement
  for multilevel graph-partitioning},'' \emph{Parallel Computing}, vol.~26,
  no.~12, pp. 1609--1634, 2000. [Online]. Available:
  \url{https://dl.acm.org/citation.cfm?id=362655}
\BIBentrySTDinterwordspacing

\bibitem{Duan2014LinearAppMWM}
\BIBentryALTinterwordspacing
R.~Duan and S.~Pettie, ``{Linear-Time Approximation for Maximum Weight
  Matching},'' \emph{J. ACM}, vol.~61, no.~1, pp. 1--23, 2014. [Online].
  Available: \url{http://dl.acm.org/citation.cfm?doid=2578041.2529989}
\BIBentrySTDinterwordspacing

\bibitem{Duan2010LinearAppMWM}
\BIBentryALTinterwordspacing
------, ``{Approximating Maximum Weight Matching in Near-Linear Time},'' in
  \emph{Proc. of the ACM/SIAM SODA 2010}.\hskip 1em plus 0.5em minus
  0.4em\relax IEEE, Oct 2010, pp. 673--682. [Online]. Available:
  \url{http://ieeexplore.ieee.org/document/5671334/}
\BIBentrySTDinterwordspacing

\bibitem{Pettie2004LinearAppMWM}
\BIBentryALTinterwordspacing
S.~Pettie and P.~Sanders, ``{A simpler linear time 2/3$-\epsilon$ approximation
  for maximum weight matching},'' \emph{Inf. Process. Lett.}, vol.~91, no.~6,
  pp. 271--276, Sept 2004. [Online]. Available:
  \url{http://linkinghub.elsevier.com/retrieve/pii/S0020019004001565}
\BIBentrySTDinterwordspacing

\bibitem{Meinel1563LinearAppMWM}
\BIBentryALTinterwordspacing
C.~Meinel and S.~Tison, ``{Linear time 1/2-approximation algorithm for maximum
  weighted matching in general graphs},'' in \emph{Proc. of the STACS
  1999}.\hskip 1em plus 0.5em minus 0.4em\relax Trier, Germany: Springer, 1999,
  pp. 259--269. [Online]. Available:
  \url{http://dl.acm.org/citation.cfm?id=1764924}
\BIBentrySTDinterwordspacing

\bibitem{Gong2017QPS}
\BIBentryALTinterwordspacing
L.~Gong, P.~Tune, L.~Liu, S.~Yang, and J.~J. Xu, ``{Queue-Proportional
  Sampling: A Better Approach to Crossbar Scheduling for Input-Queued
  Switches},'' \emph{Proc. ACM Meas. Anal. Comput. Syst. (Invention
  Disclosure-GTRC 7376)}, vol.~1, no.~1, pp. 3:1--3:33, 2017. [Online].
  Available: \url{https://dl.acm.org/citation.cfm?id=3084440}
\BIBentrySTDinterwordspacing

\bibitem{GongLiuYangEtAl2018}
L.~{Gong}, L.~{Liu}, S.~{Yang}, J.~{Xu}, Y.~{Xie}, and X.~{Wang}, ``Serenade: A
  parallel randomized algorithm suite for crossbar scheduling in input-queued
  switches,'' \emph{ArXiv e-prints (Invention Disclosure-GTRC 7648)}, Oct 2017.

\bibitem{Bayati2008BP}
\BIBentryALTinterwordspacing
M.~Bayati, D.~Shah, and M.~Sharma, ``{Max-Product for Maximum Weight Matching:
  Convergence, Correctness, and LP Duality},'' \emph{{IEEE} Trans. Inf.
  Theory}, vol.~54, no.~3, pp. 1241--1251, Mar 2008. [Online]. Available:
  \url{http://ieeexplore.ieee.org/document/4455730/}
\BIBentrySTDinterwordspacing

\bibitem{Bayati2005BP}
\BIBentryALTinterwordspacing
------, ``{Maximum weight matching via max-product belief propagation},'' in
  \emph{Proc. of the IEEE ISIT 2005}.\hskip 1em plus 0.5em minus 0.4em\relax
  Adelaide, SA, Australia: IEEE, 2005, pp. 1763--1767. [Online]. Available:
  \url{http://ieeexplore.ieee.org/document/1523648/}
\BIBentrySTDinterwordspacing

\bibitem{giaccone2003randomized}
P.~Giaccone, B.~Prabhakar, and D.~Shah, ``Randomized scheduling algorithms for
  high-aggregate bandwidth switches,'' \emph{{IEEE} J. Sel. Areas Commun.},
  vol.~21, no.~4, pp. 546--559, 2003.

\bibitem{WangMahoneyMohanEtAl2015ParallelPackingLP}
\BIBentryALTinterwordspacing
D.~Wang, M.~Mahoney, N.~Mohan, and S.~Rao, ``{Faster Parallel Solver for
  Positive Linear Programs via Dynamically-Bucketed Selective Coordinate
  Descent},'' \emph{arXiv Prepr. arXiv1511.06468}, 2015. [Online]. Available:
  \url{https://arxiv.org/pdf/1511.06468.pdf http://arxiv.org/abs/1511.06468}
\BIBentrySTDinterwordspacing

\bibitem{Vose1991AliasMethod}
M.~D. Vose, ``A linear algorithm for generating random numbers with a given
  distribution,'' \emph{{IEEE} Trans. Softw. Eng.}, vol.~17, no.~9, pp.
  972--975, Sep 1991.

\bibitem{Edelman2004ParallelPrefix}
A.~Edelman, ``Parallel prefix,''
  \url{http://courses.csail.mit.edu/18.337/2004/book/Lecture_03-Parallel_Prefix.pdf},
  2004.

\bibitem{Ladner1980PrefixSum}
\BIBentryALTinterwordspacing
R.~E. Ladner and M.~J. Fischer, ``Parallel prefix computation,'' \emph{J. ACM},
  vol.~27, no.~4, pp. 831--838, Oct 1980. [Online]. Available:
  \url{http://doi.acm.org/10.1145/322217.322232}
\BIBentrySTDinterwordspacing

\end{thebibliography}

 \appendix 
%
\section{Amortized  \texorpdfstring{$O(1)$}{O(1)} QPS Computation}\label{subsubsec: constant-time-qps}

As mentioned in Section~\ref{subsec: r-qps-serena}, 
we can perform a batch of $O(N)$ QPS computations in $O(N^2)$ time. 
In this section, we describe in details this batched QPS computation. Similar to (the standard) QPS~\cite{Gong2017QPS}, 
batched QPS also consists of two phases, namely, a batched QPS proposing phase and a batched 
QPS accepting phase.

\medskip
\noindent
{\bf Batched QSP proposing phase.} Without loss of generality, we assume 
the batch size is $m$, that is, to 
execute $m$ QPS computations in each batch. In this phase, each input port needs to 
generate $m$ QPS proposals, each of which is independently sampled proportional to the length of the 
corresponding VOQ. For ease of presentation, we only describe the batched proposing 
at input port $1$; that is identical at any other input port. Denote as 
$q_1,q_2,\cdots,q_N$ the queue 
lengths of $N$ VOQs at input port $1$ and as $q$ their total ({\it i.e.,} $q=\sum_{k=1}^{N}q_k$). 
Then, the batched QPS proposing is equivalent to generating $m$ random numbers with the given 
distribution $\mathbb{P}[\xi = j]=q_j/q, j=1,2,\cdots,N$. This can be done by Vose's algorithm~\cite{Vose1991AliasMethod} 
with $O(N)$ preprocessing time and $O(1)$ time for drawing each random number. Therefore, 
when $m = \Omega(N)$, the batched QPS proposing takes $O(m)$ time to generate $m$ proposals, that 
is $O(1)$ per proposal.

\medskip
\noindent
{\bf Batched QPS accepting.} The batched accepting phase is simply to execute the 
standard QPS accepting operation~\cite{Gong2017QPS} one after another. 
Therefore, we can perform a single QPS accepting operation for all $N$ output ports in $O(N)$ time, or $O(N)$ such operations in $O(N^2)$ time.

\section{Parallelized POPULATE}\label{subsubsec: paral-population}

As mentioned in Section~\ref{subsec: r-qps-serenade}, SERENADE~\cite{GongLiuYangEtAl2018} takes advantage of 
bit-level parallelism to reduce the practical time complexity of 
the POPULATE operation to ``$O(N)$ light''. 
However, 
this complexity in theory is still $O(N)$. 
In this section, we describe a parallel algorithm that performs POPULATE in $O(\log N)$ time using $O(N)$ processors.


Recall that, after each QPS operation, each input/output port knows whether itself is 
matched or not. The unmatched input ports and output 
ports will then be matched in a round-robin manner. That is, the first unmatched 
input port would be matched with the first unmatched output port, the second 
with the second, and so on. 

Suppose that each unmatched port (input port or output port) knows its own ranking, 
{\it i.e.,} the number of unmatched ports 
up to itself (including itself) from the first one. Then, each unmatched input 
port needs to obtain the identity of the unmatched output port with the same 
ranking. This can be done via $3$ message exchanges as follows. 
Each pair of unmatched input and output ports ``exchange'' 
their identities through a ``broker''. More precisely, 
the $j^{th}$ unmatched input port ({\it i.e.,} unmatched input port with 
ranking $j$) first sends its identity to input port $j$ ({\it i.e.,} the 
broker). 
Then, the $j^{th}$ unmatched output port also sends its identity to input port $j$ ({\it i.e.,} the 
broker). 
Finally, input port $j$ ({\it i.e.,} the 
broker) sends the identity of 
the output port with ranking $j$ to the input port (with ranking $j$). 
Thus, the input port learns the identity of the corresponding output 
port. 
Note that, since 
every pair of unmatched input port and output port 
has its unique ranking, thus they would have different ``brokers''. Therefore, all pairs can 
simultaneously exchange their messages without causing any congestion ({\it i.e.,} a port sending or 
receiving too many messages). 

It remains to parallelize the computation of ranking each port (input port or output port). 
This problem can be reduced to 
the parallel prefix sum problem~\cite{Edelman2004ParallelPrefix} as follows. Here, we 
will only show how to compute the rankings of input ports in parallel; 
that for output ports is identical. 
Let $B[1..N]$ be a bitmap that indicates whether input port $i$ is unmatched (when $B[i]=1$) or not
(when $B[i]=0$).  
For $i = 1,2,\cdots,N$, denote as $r_i$ the ranking of 
input port $i$.  It is clear that   
$r_i=\sum_{k=1}^{i}B[k]$, for any $1\le i \le N$.
In other words, the $N$ terms $r_1$, $r_2$, $\cdots$, $r_N$ are the prefix sums 
of the $N$ terms $B[1]$, $B[2]$, $\cdots$, $B[N]$. Using the Ladner-Fischer 
parallel prefix-sum algorithm~\cite{Ladner1980PrefixSum}, we can obtain 
these $N$ prefix sums $r_1$, $r_2$, $\cdots$, $r_N$ in $O(\log N)$ time (per port) using $2N$ processors (one at each input or output port).

\end{document}